\documentclass[
aps,preprint,
raggedbottom,superscriptaddress,altaffilsymbol,amsmath,amssymb]
{revtex4-1}

\usepackage{graphicx}
\usepackage{bm}
\usepackage{url}

\begin{document}

\title{Quantum Reflection above the Classical Radiation-reaction Barrier in the QED Regime}

\author{X. S. Geng}
    \affiliation{State Key Laboratory of High Field Laser Physics, Shanghai Institute of Optics and Fine Mechanics, Chinese Academy of Sciences, Shanghai 201800, China}
    \affiliation{University of Chinese Academy of Sciences, Beijing 100049, China}

\author{L. L. Ji}
    \email{jill@siom.ac.cn}
    \affiliation{State Key Laboratory of High Field Laser Physics, Shanghai Institute of Optics and Fine Mechanics, Chinese Academy of Sciences, Shanghai 201800, China}
    \affiliation{Collaborative Innovation Center of IFSA(CICIFSA), Shanghai Jiao Tong University, Shanghai 200240, China}

\author{B. F. Shen}
    \email{bfshen@mail.shcnc.ac.cn}
    \affiliation{State Key Laboratory of High Field Laser Physics, Shanghai Institute of Optics and Fine Mechanics, Chinese Academy of Sciences, Shanghai 201800, China}
    \affiliation{Collaborative Innovation Center of IFSA(CICIFSA), Shanghai Jiao Tong University, Shanghai 200240, China}
    \affiliation{Shanghai Normal University, Shanghai 200234, China}

\author{B. Feng}
    \affiliation{State Key Laboratory of High Field Laser Physics, Shanghai Institute of Optics and Fine Mechanics, Chinese Academy of Sciences, Shanghai 201800, China}
    \affiliation{University of Chinese Academy of Sciences, Beijing 100049, China}

\author{Z. Guo}
    \affiliation{State Key Laboratory of High Field Laser Physics, Shanghai Institute of Optics and Fine Mechanics, Chinese Academy of Sciences, Shanghai 201800, China}
    \affiliation{University of Chinese Academy of Sciences, Beijing 100049, China}

\author{Q. Yu}
    \affiliation{State Key Laboratory of High Field Laser Physics, Shanghai Institute of Optics and Fine Mechanics, Chinese Academy of Sciences, Shanghai 201800, China}

\author{L.G. Zhang}
    \affiliation{State Key Laboratory of High Field Laser Physics, Shanghai Institute of Optics and Fine Mechanics, Chinese Academy of Sciences, Shanghai 201800, China}
    \affiliation{University of Chinese Academy of Sciences, Beijing 100049, China}

\author{Z.Z. Xu}
    \affiliation{State Key Laboratory of High Field Laser Physics, Shanghai Institute of Optics and Fine Mechanics, Chinese Academy of Sciences, Shanghai 201800, China}
    \affiliation{University of Chinese Academy of Sciences, Beijing 100049, China}

\date{\today}

\begin{abstract}
The colliding between an ultra-intense laser pulse with a high energy electron beam is not only an important source for high-brightness gamma-rays but also a powerful approach to exploit new physics in the exotic strong-field QED regime. In the cross-colliding geometry, when radiation-reaction (RR) force is interpreted by the classical Landau-Lifshitz equation, we found that there is a distinctive barrier that allows penetration of electrons at energies beyond the barrier and blocks those of lower energies. While in the QED perspective, electrons can be well reflected (transmit) in the regime where complete transmission (reflection) is allowed classically. The reflection (transmission) is guaranteed by the quantum nature of radiation but forbidden by classical description. This effect is accompanied by the blurring of the angular distribution for scattered electrons and becomes significant for laser intensities at $2\times10^{23}\ \mathrm{W/cm^2}$ and electron energies of $\sim 10^2\ \mathrm{MeV}$; thus could be measured in the up-coming 10-100PW laser facilities. By detecting the reflection rate of the energetic electron beam after colliding and resolve the angular distribution, the results are capable of identifying the boundaries between classical and QED approaches in the strong field regime and testifying the various models describing the fundamental process.
\end{abstract}


\maketitle
Understanding the electron dynamics in relativistic laser fields has been a core interest in strong-field physics and brooded numerous key applications such as fast ignition fusion, acceleration of charged particles and producing bright X/gamma-ray sources. These advantages become strong motivations for developing the 10-100PW high-power laser systems, including SEL, ELI, XCELS, Apollon, Vulcan, SULF \cite{SEL, ELIweb, XCELSweb, Apollonweb, Vulcanweb, SULF} etc. Light intensity is likely to approach  $10^{23-24}\ \mathrm{W/cm^2}$ in the foreseeable future and promote light-matter interaction to the unprecedented radiation-dominated regime \cite{DiPiazzaRDR} or even the QED regime \cite{DiPiazzaRevModPhys,Marklund,SokolovRR}. In the new regimes, a rising interest of fundamental importance is the unique electron dynamics at extreme laser fields, in which electrons are accelerated and radiate photons of considerable energies such that recoil force is not negligible. This phenomenon is usually referred as radiation-reaction (RR).

Theoretical attempts were made to account for classical RR such as Lorentz-Abraham-Dirac equation \cite{Dirac} and Landau-Lifshitz (LL) equation \cite{LL}, both of which were derived from the assumption of continuous classical radiation. The latter is widely accepted because it resolves the nonphysical run-away solution \cite{Spohn}. Classical treatment is successive in describing accumulative RR effects. In the QED regime, stochastic radiation and high energy photon emission \cite{Pandit, SokolovModelLPI} no longer allow one to treat RR as a continuous effect and we expect to find out the role of quantum effects in RR. While photon emission and the RR force could lead to profound effects in light-matter interaction, such as electron cooling \cite{RadiationDamping, AllopticalRRSilva}, energy redistribution \cite{JillEnergyRedistribution} and anomalous trapping of electrons \cite{RRatractor, JillRRT,GonoskovART}, identifying RR in the QED regime has been obscure due to the insufficient peak laser intensities. A direct approach is to head-on collide a high intensity laser with an energetic electron beam. The laser field can be boosted by a factor of $\sim\gamma$ (electron gamma factor) in the electron rest frame so that the QED parameter $\chi=e\hbar\left|F\cdot p\right|/m^3c^4$  could reach unity \cite{Ritus}, where $F^{\mu\nu}$ is the electromagnetic tensor and $p_\mu$ is the electron four-momentum. Considerations based on this scenario have been made to observe classical RR \cite{RadiationDamping, DiPiazzaRDR,PiazzaRR,CapdessusRR} and quantum effects \cite{NeitzDiPiazzaQRR, Quenching, KeitelQRR, SilvaQRR, RidgersQRR}. Efforts were mainly focused on identifying the signature from either the radiated gamma-photons \cite{AllopticalRRSilva,KeitelQRR} or the electron dynamics \cite{Quenching,SilvaQRR}. For the latter, particularly, a quantum quenching effect is revealed in the head-on colliding geometry for few-cycle laser pulse \cite{Quenching}, by which some electrons can radiate zero energy and go through the laser field freely.

In this article, we show that new quantum features in electron dynamics arises in the cross-colliding geometry at extreme laser intensities. We noticed that classical RR could form a barrier that blocks electrons with energy below the barrier and allows electrons with energy beyond the barrier to pass. However,  in the regime where the classical RR barrier allows for full transmission (reflection), we found that an electron can be reflected (transmit) due to the quantum nature of its dynamics. In the quantum reflection (transmission) mechanism, a considerable portion of electrons get reflected (transmitted) by the laser field even when they radiate much less energies (more energies) than they do classically. This unique behavior has not been revealed in previous studies. For instance, in the quenching picture \cite{Quenching}, an electron transmits only by radiating much less energy than it does classically. In this work, we focus on the quantum-reflected electrons based on the experimental consideration that the reflected electron signal is free of the background signal from the abundant transmitted electrons.

\section{Results}
We examine the electron dynamics by perpendicularly colliding a focused laser beam to a high energy electron bunch. The tightly focused laser \cite{SalaminTightlyfocused} is polarized in $x$-direction and propagates in $z$-direction with profile of $\bm{E}=\bm{E}_0(x,y,z,w_0)\cos^2(\psi/2N)$, with $w_0$ the radius of beam waist, $\psi$ ($|\psi|<N\pi$) the phase term and $N$ the pulse length in wavelength that focuses at origin at $t=0$. We start with the simplest case where mono-energetic electrons are injected in the $y=0$ plane. A more realistic electron bunch will be discussed later. We evaluate the reflection ratio of electrons in a large parametric region.
\begin{figure}
    \centering
    \includegraphics[width=0.9\textwidth]{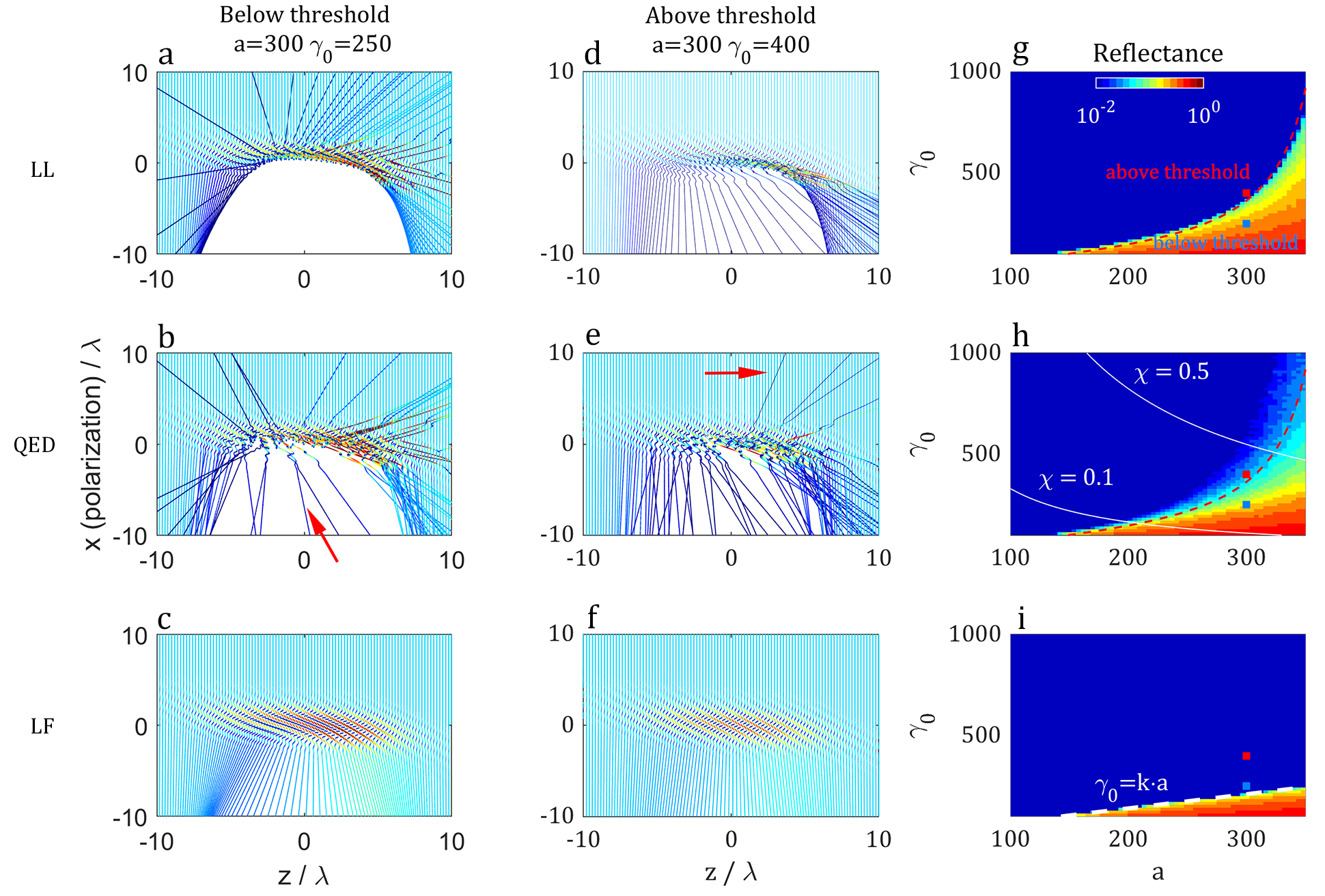}
        \caption{
                \textbf{a-f} Electron trajectories in $y=0$ plane for the LL (1st row), QED (2nd row) and LF (3rd row) cases, respectively, where $a = 300$, $\gamma_0=250$ (left), $\gamma_0=400$ (middle).
                \textbf{g-i} Electron reflectance of different ($a, \gamma_0$) in each case (LL, QED and LF), where the blue-squared dots are $(300,250)$ and red-squared dots are $(300,400)$. Reflectance maps are calculated by counting electrons  injected from $z=-10\lambda$ to $10\lambda$.
                }
        \label{fig:transmittance}
\end{figure}

\subsection{Electron reflection beyond classical RR barrier}
We compare the results between the classical approach and the QED calculation in Fig. \ref{fig:transmittance}, where $a = 300$ and $\gamma_0 = 250,400$ ($2\times10^{23}\ \mathrm{W/cm^2}$,128 $\mathrm{MeV}$,204 $\mathrm{MeV}$) are considered.
Here $\lambda=800\ \mathrm{nm}$ , $w_0=2\lambda$, pulse length is $20\lambda$ and $a = Ee\lambda/2\pi mc^2$ ($E$ is the laser electric field) is the Lorentz-invariant field strength and $\gamma_0m c^2$ is the electron initial energy. The case of pure Lorentz force (LF) is also included for full comparison. Under this parameter ($\chi<1$), the electron trajectories are greatly diverged.
When the RR effect is excluded, all electrons transmit freely through the laser beam with minor perturbation in the particle trajectories, as shown in Fig. \ref{fig:transmittance}(3rd row).
However, here is distinctive electron behavior when the classical RR (LL equation is employed) is turned on. At a fixed laser intensity of $2\times10^{23}\ \mathrm{W/cm^2}$, for $\gamma_0=250$ one sees complete reflection of electrons in the colliding vicinity (Fig. \ref{fig:transmittance}a). Increasing the electron energy to $\gamma_0=400$, we find the picture flips where all electrons transmit through the laser beam (Fig. \ref{fig:transmittance}d). The presence of such drastically different dynamic behavior by slightly varying the electron energy indicates a distinctive threshold between the two sets of parameters, where $\gamma_0=250$ and $\gamma_0=400$ correspond to below and above it respectively.
These features vanish when we switch to the  QED description. In the classical reflection ($\gamma_0 = 250$, below threshold) or transmission ($\gamma_0 =400$, above threshold) regime we see that a significant portion of electrons transmit through the laser field in the former and get reflected in the latter, as illustrated in Fig. \ref{fig:transmittance}b and \ref{fig:transmittance}e, respectively.

This disparity is universal in a large parameter range. We quantize the electron reflectance ratio in the $(a,\gamma_0)$ domain in Fig. \ref{fig:transmittance}(g-i). When RR is turned off, the only barrier for electrons to overcome is the laser ponderomotive potential. Thus an electron could always freely pass through the laser field while its initial energy dominates over the ponderomotive potential. The criterion $\gamma_0\sim a$ \cite{Heinzl_Ilderton_invariant_a} is exactly the case presented in Fig. \ref{fig:transmittance}i, where the boundary (white-dashed) is perfectly fitted by $\gamma_0=k\cdot a$. The RR effect imposes another barrier for electrons. When the laser amplitude rises, the least energy for penetration (red-dashed), namely, the barrier, grows higher than no-RR case as shown in Fig. \ref{fig:transmittance}g. Therefore one sees a clear and sharp threshold that defines reversed features in the electron reflectance, as illustrated by Fig. \ref{fig:transmittance}a and \ref{fig:transmittance}d.

However, the classical RR barrier smears when viewed in the QED perspective. In the parameter region beyond the barrier where LL allows for total transmission, some electrons still get reflected in QED, as shown by the non-zero reflectance above the red-dashed line in Fig. \ref{fig:transmittance}i. Moreover, electrons could tunnel through the beam below the barrier due to quantum behavior, which significantly lowers the reflectance as compared to the LL case.

\begin{figure}
    \centering
    \includegraphics[width=0.75\textwidth]{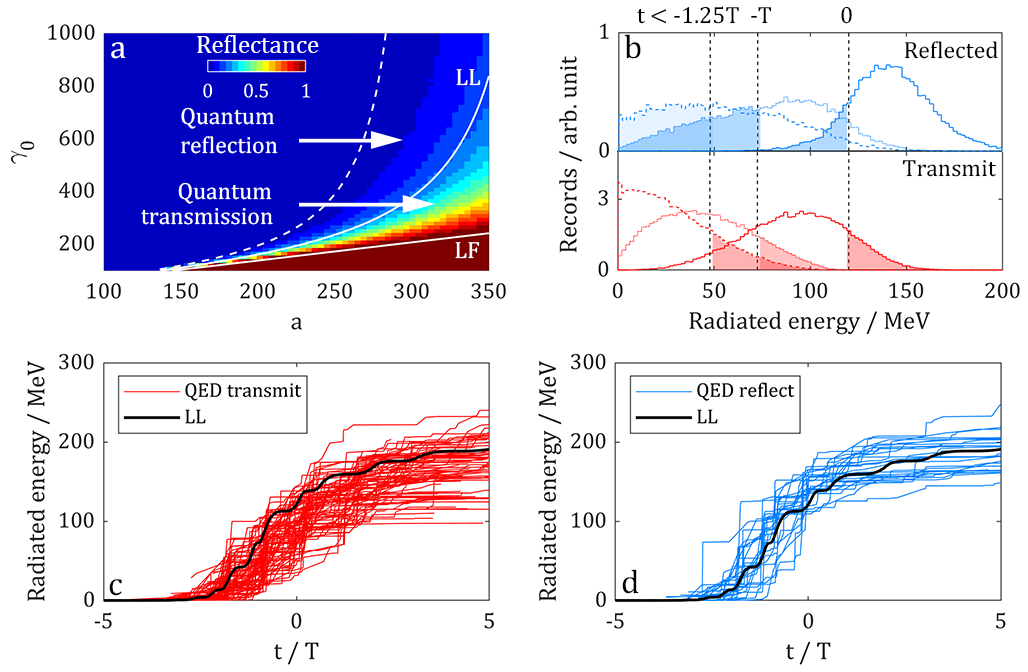}
    \caption{
            \textbf{a} Electron reflection probability from QED-MC calculation (colormap) when initial injecting position is fixed at $x_0=c t_0, z_0=0$ at $t=-t_0$; the white lines are the threshold for classical LL and LF modelling; dashed line denotes the 0-reflectance threshold for QED.
            \textbf{b} Histogram of energy loss in QED-MC during $t<-1.25T$(left), $t<-T$(middle),  $t<0$(right) for the reflected (blue) and transmitted (red) electrons measured at ($a_0=300,\gamma_0=400$), where $T=\lambda/c$. The energy loss of LL (black-dashed) are single-valued.
            \textbf{c-d}, Energy loss of the transmitted (red) and reflected (blue) electrons in QED and in LL (black).
            }
    \label{fig:energyLoss}
\end{figure}

\subsection{QED effects: anomalous reflection}
The QED effect can be best understood by looking at one electron injected at one fixed position for multiple times. The least penetration energy for classical LF and LL (solid lines) is definite as shown in in Fig. \ref{fig:energyLoss}a. These two curves coincide with the many-particle modelling in Fig. \ref{fig:transmittance}g and \ref{fig:transmittance}i.
In the QED picture, we repeat the colliding process for $10^4$ times at each set of $(\gamma_0, a)$ where the reflectance probability is defined by $\frac{\text{reflected}}{\text{total}}$. The reproduced probability map in Fig. \ref{fig:energyLoss}a shows three distinctive regimes. Below the LF boundary is the forbidden zone where the initial electron energy is too small to overcome the ponderomotive barrier.
The LL boundary indicates determined electron dynamic in the classical picture, which reveals the threshold for transmission and reflection shown in Fig. \ref{fig:transmittance}a and \ref{fig:transmittance}d. While in the QED picture, electron dynamic is stochastic and the reflectance is smeared near the LL boundary.
The quantum transmission region lies between the LL and the LF boundaries, which becomes more prominent as $\chi$ increases. Above the LL boundary is above-threshold zone where electron energy dominates the classical barrier.
The non-zero reflectance above LL boundary in QED indicates the quantum reflection when total transmission is allowed by classical LL. It should be noticed that the LL boundary can be modified by  quantum correction \cite{BaierQuantumCorr}. However, these quantum effects can always happen due to the stochastic behavior.

We further count the radiated energy for each electron in Fig. \ref{fig:energyLoss}b for $a=300,\gamma_0=400$ during $t<0$, as simulation starts at $t=-t_0$ and electron is initialized at $x_0=c t_0, z_0=0$ outside the laser. The LL equation gives a definite single-valued  radiated energy at each interaction time. In contrast, energy loss in the QED modeling deviates in several ways.
First, QED-MC exhibits broadened distribution ranging from 0 to 200 MeV due to random radiation, meaning that the electron could lose all the initial energy or radiate nothing.
Second, the averaged radiated energy is higher for reflected electrons and lower for transmitted electrons, which is similar to the quantum quenching effect \cite{Quenching}.

The most interesting feature in Fig. \ref{fig:energyLoss}b is that  some electrons may radiate more (less) than classical calculation but still transmit (get reflected) as shown by the colored area. These anomalously transmitted/reflected electrons are presented at each colliding phase $t=-1.25T$, $-1T$, and $0$, as shown by the records in Fig. \ref{fig:energyLoss}b. The mechanism is further confirmed in Fig. \ref{fig:energyLoss}c and \ref{fig:energyLoss}d by tracking the energy loss of each electron trajectory during collision. Therefore, a new QED feature is explicitly shown here, which has not been revealed previously.  The fact that an electron does not necessarily radiate more energy to be reflected or less to transmit is a most profound reflection of the stochastic quantum laws in this regime that is forbidden by any classical model.

\subsection{QED effects-angular feature}
We consider at $a=350$ a more realistic electron bunch of spatial size $\Delta y=\Delta z=100\lambda$ at FWHM of super-Gaussian profile, $\Delta x=4\lambda$ at FWHM of Gaussian profile with $E=102$ MeV, 5\% energy spread, 10 mrad angular divergence \cite{Liujs,Thomson_bunch} and peak density of $\sim 10^{15} \text{cm}^{-3}$. The relatively large transverse size can significantly lower down the aiming difficulty in the cross-colliding geometry. The collective plasma field can be neglected due to the very low electron density such that test particle simulation is sufficient. Otherwise  particle-in-cell (PIC) simulations are required.

\begin{figure}
    \centering
    \includegraphics[width=0.5\textwidth]{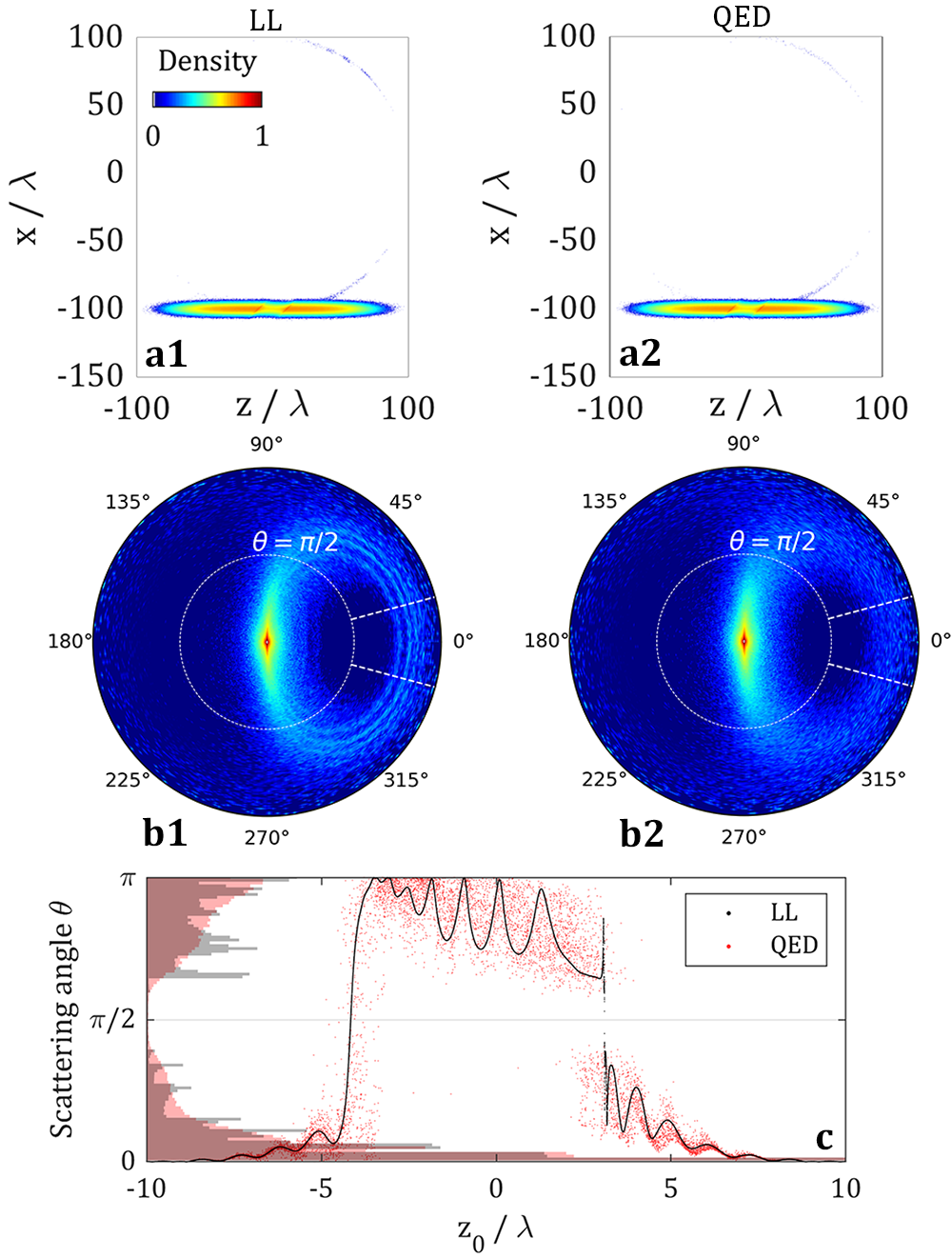}
    \caption{
            \textbf{a} Electron density distribution (in log) after cross-colliding with laser for LL (a1) and QED (a2).
            \textbf{b} Angular distribution of electrons after collision for LL (b1) and QED (b2). $\theta$ is polar angle to $-x$ axis and $\phi$ the azimuthal angle to $z$ axis.
            \textbf{c} Scattering angle $\theta$ of test particles initially injected at different phases (or $z_0$) in the $y=0$ plane, similar to Fig. \ref{fig:transmittance}. Horizontal bars are electron number records along $\theta$.
            }
    \label{fig:angle}
\end{figure}

The results are presented in Fig. \ref{fig:angle}a. Laser pulse drills through the bunch and scatters electrons to large angles. Therefore one can focus on the scattered electrons free from the disturbance of background electrons in the bunch. For long distance propagation, it is very important to identify the angular distribution after collision. In Fig.  \ref{fig:angle}b1, we see clear periodic structures in the LL case, where $\theta$ is the polar angle to $-x$ and $\phi$ the azimuthal angle to $z$; when we switch to the QED case, such structures vanish in Fig. \ref{fig:angle}b2.
In the classical case, there is a specific correlation between the scattering angle and injected phase, as shown in Fig. \ref{fig:angle}c. Thus one observes the oscillation along scattering angle $\theta$ (black horizontal bars), which represents the structure of the laser field. However, in the QED case, scattering angle is no longer single-valued to the injecting phase. This is induced by stochastic effect that allows electrons to enter phases forbidden by the classical model. The broadened scattering angle in QED thus smears the oscillating structure presented in the classical case.

\section{Experimental consideration}
\begin{figure}
    \centering
    \includegraphics[width=0.75\textwidth]{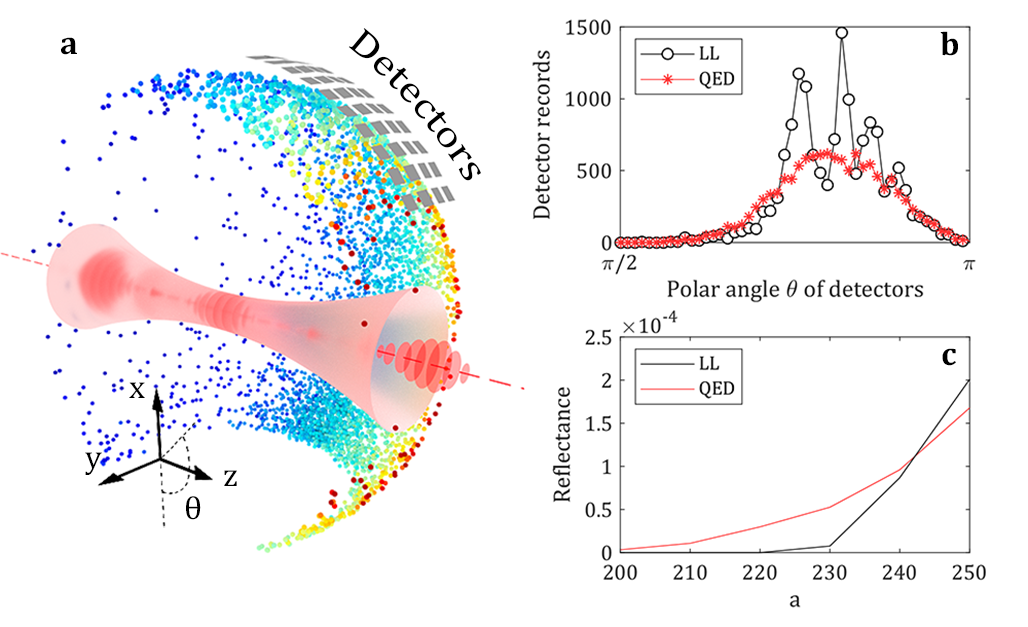}
    \caption{\label{fig:experiment}
    \textbf{a} Illustration of scattered electrons and detectors. Detector array is placed in the reflection direction to capture the reflected electrons.
    \textbf{b} Recorded electron number of detectors along $\theta$.
    \textbf{c} Reflectance of all electrons for different field strength.
    }
\end{figure}

One can leverage the new feature to experimentally probe the QED nature at extreme
intensities. In this work, we focus on the quantum-reflected electrons as the reflected electron signal is free of the background signal from the abundant transmitted electrons.
These novel effects can be captured experimentally by distributing electron number detectors along $\theta$ and $\phi$ angles and recording the scattered electrons, as shown in Fig \ref{fig:experiment}a. The records, accumulated from $\phi=-15^\circ$ to $15^\circ$ (sections between dashed lines in Fig. \ref{fig:angle}b) as a function of $\theta$ (50 sets of detectors along $\theta$ ranging from $\pi/2$ to $\pi$) is shown Fig. \ref{fig:experiment}b. Each detector thus corresponds to an accepting area of $\sim 6\text{cm}\times6\text{cm}$ for 2 meter propagation distance. Clear oscillations (spikes and valleys) are reproduced in the detectors along $\theta$ in the classical case. This feature vanishes in the QED picture. Therefore, the absence of such oscillation structure in angular distribution is an explicit evidence of the QED process.
We integrate electron numbers in $\pi/2<\theta<\pi$ and obtain the total reflectance at different field strengths in Fig. \ref{fig:experiment}c. The reflection curves show a sharp transition at $a=220$ for LL which exactly reflects the classical RR barrier. For QED, however, the growth is smoothed due to quantum mechanism. This is consistent with the quantum reflection region in Fig. \ref{fig:energyLoss}a. We thus conclude that the measurement of electron reflectance along with the angular structure will provide a clear proof of quantum RR.

\section{Discussion}
\subsection{Pair-production}
\begin{figure}
    \includegraphics[width=0.5\textwidth]{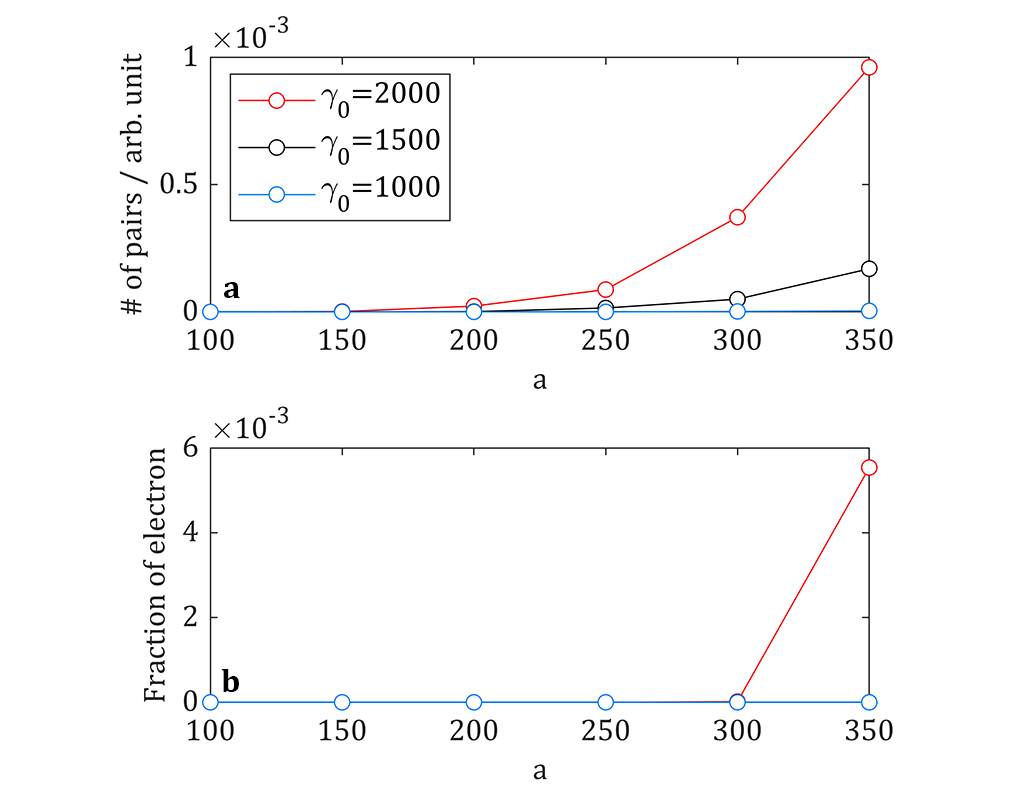}
    \caption{
            \textbf{a} Electron-positron pairs produced in the collision.
            \textbf{b} The fraction of pair-produced electrons for the reflected electrons.}
    \label{fig:pair-production}
\end{figure}

Breit-Wheeler pair-production \cite{BreitWheelerPP} in the cross-collision geometry is investigated in a large parameter range with the PIC code SMILEI \cite{smilei}. We use the same laser and electron bunch configuration as that in Fig. \ref{fig:angle} but in 2-dimensional form. The cell size in the PIC simulation is $0.04\lambda\times0.04\lambda$ with 10 particles per cell. Results are presented in Fig. \ref{fig:pair-production}, where the pair-production is strongly suppressed by $\chi\sim\gamma\cdot a$. The fraction of the pair-produced electron is also small when compared to all the reflected electrons. Therefore, in the domain of interest ($\chi<1$), pair-production does not affect the phenomenon of quantum reflection.

\subsection{Comparison with PIC results}
\begin{figure}
    \centering
    \includegraphics[width=0.75\textwidth]{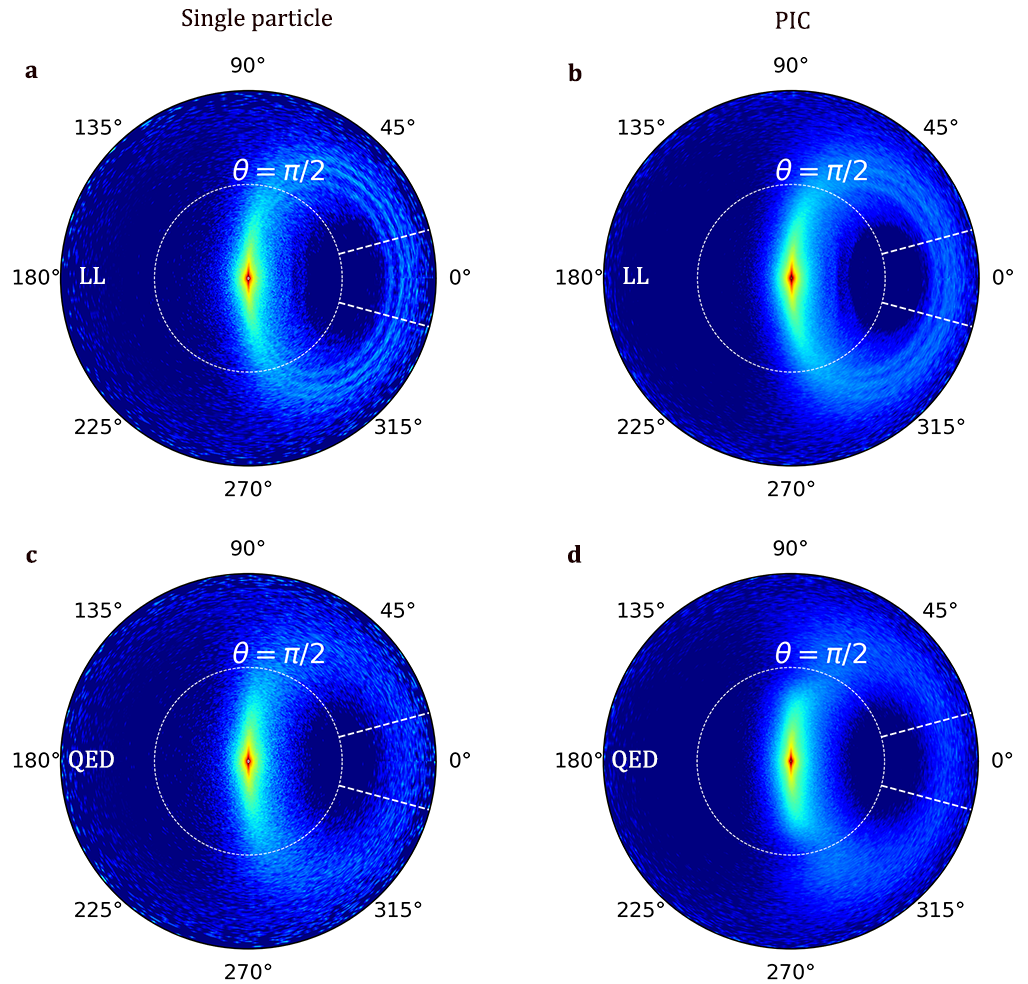}
    \caption{
            Angular distribution after cross-collision for single-particle simulation (a,c) and PIC simulation (b,d), where \textbf{a,b} are the results of LL equation and \textbf{c,d} are the results of QED radiation.
            }
    \label{fig:comparision}
\end{figure}

All the results are calculated with our single-particle model. Here we compare our results of angular distribution in Fig. \ref{fig:angle} with PIC simulation, where the laser and electron bunch configurations in the PIC simulation are the same with the single-particle simulation. The cell size in the PIC simulation is $0.04\lambda\times0.08\lambda\times0.08\lambda$ with 1 particle per cell, where the first dimension is the laser propagation direction. Comparison of the results are shown in Fig. \ref{fig:comparision}. Although the single-particle simulation ignores the electrons' interaction and PIC simulation introduces artificial space-grid, the consistency with each other still shows fidelity of our simulations.

\section{Conclusion}
In conclusion, we find that perpendicular laser-electron collision is a unique approach to distinguish the classical and quantum dynamics in the strong field regime. By calculating the reflectance of injected electrons, we revealed the classical RR-barrier threshold that can be penetrated by electrons of energy above the threshold and blocks those below. This classical barrier grows to infinity that blocks electron of arbitrarily high energy when field strength reaches a critical value. While in QED, electron may get reflected even when electron energy dominates the classical barrier and transmit the barrier when electrons energy is below the barrier. We attribute the effect to stochastic radiation and consequent nonlinear motion and pointed out the parameter region for quantum reflection. For experimental consideration, one can resolve the angular distribution of reflected electron in a cross-collision to identify quantum RR effects.

We notice that recent experimental efforts have been paid to collide laser-wakefield accelerated electrons with counter-propagating, high intensity laser pulses to produce X/gamma-rays via Compton scattering \cite{gamma_Chen,gamma_Sarri,gamma_Yan} and to create RR events\cite{RR_Cole,RRexpDiPiazza}. In the former, RR was not active due to relatively low laser intensities. In experiment by Cole et al., the collision probability  was limited by the techniques of time synchronization and spatial overlap. Thus successful events were necessarily identified through extensive theoretical modeling and comparison to experimental results\cite{RR_Cole}. In experiment by Poder et al., a more precise model beyond constant cross field approximation is required to account for the results\cite{RRexpDiPiazza}. Wistisen et al. \cite{RR_wistisen} measures photon spectrum in a positron-crystal collision but did not find a valid theory that fully accounts for the results. We believe the new QED features of electron dynamics revealed here can be explicitly probed by the measurement proposed in Fig. \ref{fig:experiment}. The method is in principle independent of the theoretical modeling by double measuring the angular structure and the reflectance.

\section{Methods}
\begin{figure}
    \includegraphics[width=0.75\textwidth]{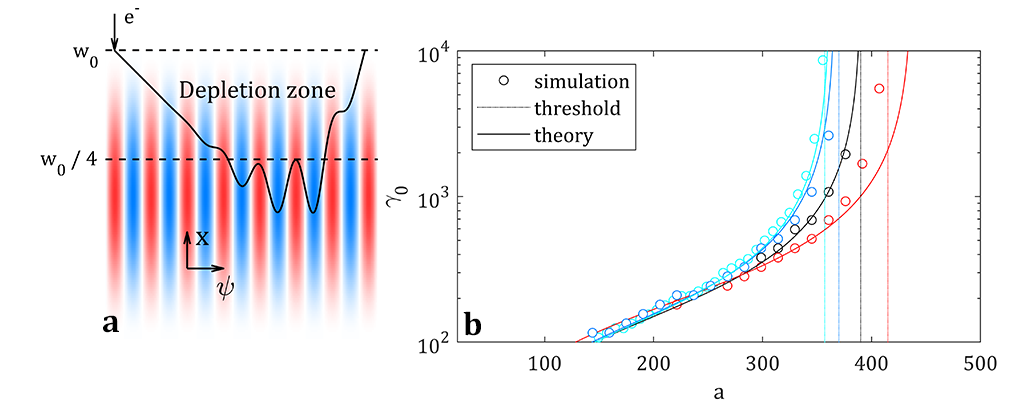}
    \caption{
            \textbf{a} Electron trajectory in the $\psi-x$ space in a plane laser pulse with a $cos^2$ beam waist $w_0=4\lambda$ and peak amplitude $a=400$. The electron is initially positioned at $z=0$ with $\gamma_0=10^4$.
            \textbf{b} The boundary curve from LL transmittance map (circle) and corresponding results of Eq. (\ref{eq:leastgamma0}) (solid), for different beam size $w_0$ (cyan $3\lambda$, blue $2.5\lambda$, black $2\lambda$, red $1.5\lambda$) of Gaussian profile. Dashed lines denote the threshold of the classical LL barrier, namely $a_{cr}$.
            }
    \label{fig:rapidexhaustion}
\end{figure}

In classical picture, we solve the equation of motion $d\bm{p}/dt=\bm{F}_L+\bm{F}_{RR}$ numerically for each electron and track their trajectories. Here, $\bm{F}_L$ is the Lorentz force and $\bm{F}_{RR}$ is the RR force taking the dominating term of LL equation\cite{LL}. In the classical framework, the electron dynamics is deterministic in the sense that it always experiences the same amount of damping if the injection position is constant. We consider an electron initially located at the upper edge of the laser field, as seen in Fig. \ref{fig:rapidexhaustion}a. The trajectory is presented in the $\psi-x$ space, where $\psi=\omega t - kz$. One notices that before the trajectory turns staggered, there is a depletion zone where it loses the kinetic energy rapidly along a straight line in the space-time domain. This process, namely \textit{rapid exhaustion}, is the key making electron possible to be reflected by laser field. For electrons with energy far beyond ponderomotive potential, they must lower down their momenta to a certain level through the rapid exhaustion phase in RR to be reflected by the laser ponderomotive potential.

We simplify the geometry to one-dimensional (1D), since the electron moves in the polarization plane and maintains a straight trajectory. To give an explicit formula, the laser is approximated to a plane wave
\begin{equation*}
    E_x=E_0 \cos(kz-\omega t)\cos^2\left(\pi x/2w_0\right)
\end{equation*}
with beam waist $w_0$.
Since RR is dominant in the depletion zone, we only consider RR force
\begin{equation*}
    \bm{F}_{RR}\approx -\frac{2e^4}{3m^2c^4}\gamma^2\bm{\beta}
    [\left(\bm{E}+\bm{\beta} \times \bm{H}\right)^2-(\bm{E}\cdot\bm{\beta})^2]
\end{equation*}
where we only take the dominant term proportional to $\gamma^2$. We have the rate of energy loss from classical radiation $dE/dt=\bm{F}_{RR}\cdot\bm{v}$, where $v_x\approx-c$. Along the straight trajectory one has $t=(w_0-x)/c$ then $dE/dt$ only depends on $x$. Then electron energy evolves as
\begin{equation*}
    m c^2\frac{d\gamma}{dt}= m c^2\frac{d\gamma}{dx}\frac{1}{-c}
    =-\frac{2e^4}{3m^2c^3}\gamma^2E_x^2(x)\beta_x^2
\end{equation*}
where $\beta_x^2\approx1$. Then we have
\begin{equation*}
    \int_{\gamma_0}^{\gamma(x)} \gamma^{-2} d\gamma =
    \int_{w_0}^{x} \frac{2e^4}{3m^3c^4} E_{x}^2(x) dx
\end{equation*}
Finally we obtain the evolution of $\gamma(x)$ along $x$
\begin{equation*}
    \gamma\left(x\right)=\frac{1}{1/\gamma_0 + a^2 \mathbb{I}_{w_0}\left(x\right)}
\end{equation*}
where $\mathbb{I}_{w_0}(x)$ is an integral relevant to laser profile. It weakly depends on initial $z_0$ thus we take $z_0 = 0$ for convenience. From Fig. \ref{fig:rapidexhaustion} we see that the depletion zone ends at about $w_0/4$ away from the laser axis, thus we evaluate the integration $\mathbb{I}_{w_0}\left(x\right)$ from $x = w_0$ to $w_0/4$. The least energy for an electron to penetrate ponderomotive potential is approximated by demanding $\gamma(w_0/4)\sim k\cdot a$, from which we can determine the least energy ($\gamma_0m c ^2$) to penetrate the laser beam considering strong RR
\begin{equation}\label{eq:leastgamma0}
    \gamma_0=\frac{1}{1/ka - a^2\mathbb{I}_{w_0}(w_0/4)}
\end{equation}
where $k$ can be determined by the boundary (white-dashed) in Fig. \ref{fig:transmittance}f. Eq. (\ref{eq:leastgamma0}) depicts the boundary of the classical LL transmittance map. For comparison, we numerically calculate the LL boundary of transmittance in Fig. \ref{fig:energyLoss}a and compare it with Eq. (\ref{eq:leastgamma0}) in Fig. \ref{fig:rapidexhaustion}b. One sees excellent agreement between the two. For comparison, the $w_0$ of the $cos^2$  profile is scaled to fit the Gaussian profile of the laser in our simulation ($w_{0,cos^2}=1.71w_{0,Gaussian}$). Eq. (\ref{eq:leastgamma0}) reveals the unique LL threshold by setting the denominator to null $1/k a = a^2\mathbb{I}_{w_0}(w_0/4)$, suggesting an infinite value for the initial electron energy. The critical field strength then scales as
\begin{equation}\label{eq:a_cr}
    a_{cr,w_0}=[k\cdot \mathbb{I}_{w_0}(w_0/4)]^{-1/3}\sim w_0^{-1/3}
\end{equation}
where $\mathbb{I}_{w_0}(x)$ is basically linear to $w_0$. Eq. (\ref{eq:a_cr}) reveals the important fact that the LL barrier threshold does not depend on the initial momentum of the colliding electrons. It is only a function of the laser profile. We notice that the scaling of Eq. (\ref{eq:a_cr}) is consistent with the criteria by which RR dominates over LF \cite{FedotovRRvsPondero, BulanovRDR} and radiation trapping happens \cite{JillRRT, GonoskovART}. Particularly, we see that from Eq. (\ref{eq:a_cr}) that although relevant, the barrier dependence on the laser beam size is relatively weak. Therefore it is favored by tight focusing of finite laser peak powers, as $a\sim w_0^{-2}$ while $a_{cr,w_0}\sim w_0^{-1/3}$.

Quantum radiation is treated stochastically while the motion between the emission events is classical. One can do so because the de Broglie wavelength of an ultra-relativistic electron is much smaller than the optical laser wavelength. The emission rate in the QED regime is given in Ref. \cite{QED,Ritus}. We use the synchrotron radiation configuration \cite{Erber,SokolovTernov} based on the solution of a `dressed' electron in external fields; the photon emission is modeled by the synchrotron spectrum \cite{Ternov1995}
\begin{equation*}
    F(\chi,\delta) = (1-\delta)
    \left[y\int_y^\infty dy' K_{5/3}(y')+\frac{\delta^2}{1-\delta}y K_{2/3}(y) \right]
\end{equation*}
where $y=\frac{2}{3}\chi^{-1}\frac{\delta}{1-\delta}$ and $\delta=|F\cdot\hbar k|/|F\cdot p|\approx\hbar\omega/\gamma m c^2$ represents the photon energy normalized by the electron energy; $\hbar k_\mu$ is four-momentum of the photon. Photon emission is triggered by a modified event generator \cite{PICQED} that resolves the cutoff issue in the low energy region of the radiation model. In the QED-MC (Monte-Carlo) algorithm, emission probability and photon energy are determined by two independent random number $r_1$ and $r_2$. If $r_2<P(r_1)$, a photon of $\hbar\omega=r_2\gamma m c^2$ is emitted, where $P(r)$ is the probability density function constructed from $F(\chi,\delta)$.

\noindent\textbf{Acknowledgments} This work is supported by the National Science Foundation of China (Nos. 11374317 and 11335013) and the Strategic Priority Research Program of Chinese Academy of Sciences (Grant No. XDB16000000).

\noindent\textbf{Author contribution}  L. L. J.  and X. S. G. proposed the research. X. S. G. did the numerical modeling and simulations under supervision by L. L. J. and B. F. S.;  X. S. G and L. L. J. prepared the manuscript, with suggestion from B. F. S., B. F., Z. G., Q. Y., L.G. Z.,  and Z. Z. X.

\noindent\textbf{Additional information}

\noindent\textbf{Competing interests} The authors declare no competing interests.

\end{document}